\documentstyle[12pt]{article}
\newcommand{\bce}{\begin{center}}
\newcommand{\ece}{\end{center}}
\newcommand{\beq}{\begin{equation}}
\newcommand{\eeq}{\end{equation}}
\newcommand{\bea}{\vspace{0.25cm}\begin{eqnarray}}
\newcommand{\eea}{\end{eqnarray}}

\newcommand{\ba}{\begin{array}}
\newcommand{\ea}{\end{array}}

\newcommand{\doublespace}{
    \renewcommand{\baselinestretch}{1.6}\large\normalsize}

\def\lsim{\mathrel{\rlap{\lower4pt\hbox{\hskip1pt$\sim$}}
    \raise1pt\hbox{$<$}}}	  %less than or approx. symbol
\def\gsim{\mathrel{\rlap{\lower4pt\hbox{\hskip1pt$\sim$}}
    \raise1pt\hbox{$>$}}}	  %greater than or approx. symbol

\def\Pom{{\bf I\!P}}

\def\lsim{\mathrel{\rlap{\lower4pt\hbox{\hskip1pt$\sim$}}
    \raise1pt\hbox{$<$}}}         %less than or approx. symbol
\def\gsim{\mathrel{\rlap{\lower4pt\hbox{\hskip1pt$\sim$}}
    \raise1pt\hbox{$>$}}}         %greater than or approx. symbol

\def\Pom{{\bf I\!P}}

  \advance\oddsidemargin  by -1in
  \advance\evensidemargin by -1in
  \advance\topmargin      by -0.5in
\def\lsim{\mathrel{\rlap{\lower4pt\hbox{\hskip1pt$\sim$}}
    \raise1pt\hbox{$<$}}}         %less than or approx. symbol
\def\gsim{\mathrel{\rlap{\lower4pt\hbox{\hskip1pt$\sim$}}
    \raise1pt\hbox{$>$}}}         %greater than or approx. symbol

\def\Pom{{\bf I\!P}}
\def\beq{\begin{equation}}
\def\endeq{\end{equation}}
\def\arr{\begin{eqnarray}}
\def\endarr{\end{eqnarray}}
\makeindex

\doublespace

\begin{document}

%\large

%{ \huge The last update:  16 January 1993\vspace{2.0cm}\\}
 \phantom{.}{\bf \Large \hspace{10.0cm} KFA-IKP(Th)-1994-5 \\
 \phantom{.}\hspace{11.9cm}16 January   1994\vspace{0.4cm}\\ }
%}

\begin{center}
{\bf\sl \huge BFKL evolution and universal structure function
at very small $x$.}
\vspace{0.4cm}\\
{\bf \large
N.N.~Nikolaev$^{a,b}$, and B.G.~Zakharov$^{b}$
\bigskip\\}
{\it
$^{a}$IKP(Theorie), KFA J{\"u}lich, 5170 J{\"u}lich, Germany
\medskip\\
$^{b}$L. D. Landau Institute for Theoretical Physics, GSP-1,
117940, \\
ul. Kosygina 2, Moscow 117334, Russia.
\vspace{1.0cm}\\ }
{\Large
Abstract}\\
\end{center}
The Balitskii-Fadin-Kuraev-Lipatov (BFKL) and
the Gribov-Lipatov-Dokshitzer-Altarelli-Parisi (GLDAP)
evolution equations for the diffractive deep inelastic
scattering at ${1\over x} \gg 1$
are shown to have a common solution in the weak coupling
limit:
$F_{2}(x,Q^{2})\propto [\alpha_{S}(Q^{2})]^{-\gamma}
\left({1\over x}\right)^{\Delta_{\Pom}}$.
The exponent $\gamma$ and the pomeron
intercept $\Delta_{\Pom}$ are related by $\gamma\Delta_{\Pom}=
{4\over 3}$ for the $N_{f}=3$ active flavors. The existence of this
solution implies that there is no real clash between the
BFKL and GLDAP descriptions at very small $x$.
We present derivation
of this solution in the framework of our generalized BFKL
equation for the dipole cross section,
discuss conditions
for the onset of the universal scaling violations and analyse
the pattern of transition from the conventional
Double-Leading-Logarithm approximation for the GLDAP evolution to
the BFKL evolution at large ${1\over x}$.
 \bigskip\\

\begin{center}
E-mail: kph154@zam001.zam.kfa-juelich.de
\bigskip\\
{ Submitted to: \sl Physics Letters B }
\end{center}

%\doublespace
\pagebreak

%-----------------

%                 Section  1

%-----------------

\section{ Introduction.}

%-----------------

The asymptotic behavior of diffractive scattering in perturbative
QCD is usually discussed in the framework of the BFKL pomeron [1].
Recently, we took advantage of exact diagonalization of
the $S$-matrix of diffractive scattering in terms of the dipole
cross section [2,3] and derived the generalized BFKL equation for
the perturbative dipole cross section [4-6]. In
[4-6] we studied the spectrum of our generalized BFKL
equation and determined the
intercept of the pomeron and the asymptotic dipole cross section
for the rightmost singularity in the complex $j$-plane.

The central result of the present paper is a derivation of the
asymptotic pomeron solution of our generalized BFKL equation,
\beq
\sigma_{\Pom}(\xi,r)=\sigma_{\Pom}(r)\exp(\Delta_{\Pom}\xi)
\propto r^{2}
\left[{1\over \alpha_{S}(r)}\right]^{\gamma-1}\exp(\Delta_{\Pom}\xi)\, ,
\label{eq:1.1}
\endeq
where the exponent $\gamma$ is related to the pomeron intercept
$\Delta_{\Pom}$ by
\beq
\gamma={12\over \beta_{0}\Delta_{\Pom}} \, .
\label{eq:1.2}
\endeq
Here $\beta_{0}=11-{2\over 3}N_{f}=9$ for $N_{f}=3$
active flavours,
$r$ is the transverse size of the colour dipole,
$\alpha_{S}(r)$ is the running QCD coupling and
$\xi=\log({x_{0}\over x})$, where $x_{0}\sim $0.1-0.01 corresponds
to the onset of the leading-$\log({1\over x})$ approximation.
We shall demonstrate that, to an accuracy $\sim \alpha_{S}(r)$,
the new solution (\ref{eq:1.1}) is a low-$x$ limit of solutions
of the conventional GLDAP equation [7]. This new
solution must be contrasted to: \begin{itemize}
\item[i)]
 The
Double-Leading-Logarithm Approximation (DLLA) solution [4] to the
GLDAP evolution equation,
\beq
\sigma_{DLLA}(\xi,r)=\sigma(0,r)\sum_{n=0}{\eta^{n} \over n!(n+1)!}
\sim r^{2}\alpha_{S}(r)\log\left[{1\over \alpha_{S}(r)}\right]
{1 \over \sqrt{\eta}}\exp(2\sqrt{\eta})\, ,
\label{eq:1.3}
\endeq
where $\eta$ is the expansion parameter of DLLA,
$
\eta ={12\over \beta_{0}}\xi L(r)$ and $L(r)=
\log\left[{1\over \alpha_{S}(r)}\right]$.
\item[ii)]
The solution of the scaling BFKL equation [1]
in the case of fixed $\alpha_{S}$,
\beq
\sigma_{\Pom}(\xi,r)
\propto r\exp(\Delta_{\Pom}\xi) \, .
\label{eq:1.5}
\endeq
\end{itemize}
The conventional DLLA solution (\ref{eq:1.3}) summs the leading powers
$\xi^{n}L(r)^{n}$, our new solution (\ref{eq:1.1}) manifestly
summs all powers of $L(r)$.

The long going debate in the literature on the clash between,
and transition from,
the GLDAP to the BFKL regime with rising ${1\over x}$, is centered
on a comparison of the DLLA solution (\ref{eq:1.3}) and the
scaling BFKL solution (\ref{eq:1.5}) (for the recent review
and references see [8,9]). The existence of our new
solution of the generalized BFKL equation has the two major
implications: Firstly, there is no real clash
between the GLDAP and BFKL evolutions and the GLDAP
evolution remains a viable description of deep
inelastic scatering at very large ${1\over x}$ in the perturbative
regime of $\alpha_{S}(Q^{2})\ll 1$. Secondly, comparison of
solutions (\ref{eq:1.1}) and (\ref{eq:1.5}) shows the dramatic
difference between the cases of the running and fixed strong
coupling for the small-$x$ behavior (see also below, a discussion of
the so-called DLLA identity).
We conclude this introduction citing the universal structure
function
\beq
F_{2}(x,Q^{2})\propto
\left[{1\over \alpha_{S}(Q^{2})}\right]^{\gamma}
\left({1\over x}\right)^{\Delta_{\Pom}}  \, ,
\label{eq:1.6}
\endeq
which follows from the dipole cross section (\ref{eq:1.1}).
Most unfortunately, the onset of this
universal regime only is expected
beyond the kinematical range of the HERA experiments.

The paper is organized as follows: We begin with the brief review of
the lightcone description of deep inelastic scattering in terms
of the dipole cross section and present our generalized BFKL
equation for the dipole cross section. Then we discuss
a transition from from the DLLA solution to GLDAP equation
to the BFKL solution. We introduce a new comparison of the GLDAP and
BFKL solutions in terms of the so-called DLLA identity, and present
a derivation of our new solution (\ref{eq:1.1}) and of
the universal scaling violations at asymptotically large
${1\over x}$. Finally, we comment on the significance
of the consistent use of the running coupling for the pomeron
cross section. In the conclusions section we summarize our main
results.

%-------------------------------------------------

%-------------------------     Section 2

%\section{Deep inelastic scattering in the dipole-cross section
%representation and the generalized BFKL equation}

In the region of very large ${1\over x}$ the virtual photoabsorption
can be viewed as interaction with the target proton (nucleus) of
the multipartonic lightcone Fock states
($q\bar{q},\, q\bar{q}g...$) of the
photon, which are formed at large distance $\Delta z \sim
{1\over m_{p}x}$ upstream the target. Consequently, the longitudinal
momentum partitions $z_{i}$ and the transverse separations
$\vec{\rho}_{i}$ of
partons are conserved in the scattering process, and the diffractive
$S$-matrix is exactly diagonalized in the $(\vec{\rho},z)$
representation [2]. Photons do not couple to gluons and the
higher $q\bar{q}g_{1}...g_{n}$ states are
radiatively generated from the $q\bar{q}$ dipole of size $\vec{r}$.
In [4] we gave regular procedure for construction
of the corresponding multiparton
lightcone wave functions
and of the multiparton total cross sections.
The Fock states with $n$ gluons give the
$\propto \log^{n}({1\over x})$ contribution to the total
photoabsorption cross section, which can be reabsorbed into the
energy dependent dipole cross section
\beq
\sigma(\xi,r)=\sum_{n=0}{1\over n!}\sigma_{n}(r)
\xi^{n}\, .
\label{eq:2.1}
\endeq
The total photoabsorption cross section can be written as an
expectation value [2]
\beq
\sigma_{T,L}(\gamma^{*}N,\xi,Q^{2})=
\int_{0}^{1} dz\int d^{2}\vec{r}\,\,
\vert\Psi_{T,L}(z,r)\vert^{2}\sigma(\xi,r)
\label{eq:2.2}
\endeq
over the wave function of the $q\bar{q}$ Fock state.
The wave functions of the (T) transverse
and (L) longitudinal virtual
photon of virtuality $Q^{2}$ were derived in [2] and read
\beq
\vert\Psi_{T}(z,r)\vert^{2}={6\alpha_{em} \over (2\pi)^{2}}
\sum_{1}^{N_{f}}Z_{f}^{2}
\{[z^{2}+(1-z)^{2}]\varepsilon^{2}K_{1}(\varepsilon r)^{2}+
m_{f}^{2}K_{0}(\varepsilon r)^{2}\}\,\,,
\label{eq:2.3}
\endeq
\beq
\vert\Psi_{L}(z,r)\vert^{2}={6\alpha_{em} \over (2\pi)^{2}}
\sum_{1}^{N_{f}}4Z_{f}^{2}\,\,
Q^{2}\,z^{2}(1-z)^{2}K_{0}(\varepsilon r)^{2}\,\,,
\label{eq:2.4}
\endeq
where $K_{1}(x)$ is the modified Bessel function,
$\varepsilon^{2}=z(1-z)Q^{2}+m_{f}^{2}$,
$m_{f}$ is the quark mass and $z$ is
the fraction of photon's light-cone
momentum carried by one of the quarks of the $q\bar{q}$ pair
($0 <z<1$). Then, the structure function is calculated as
$ F_{2}(\xi,Q^{2})=Q^{2}
\left[\sigma_{T}+\sigma_{L}\right]/(4\pi^{2}\alpha_{em})$.
Notice that the dipole cross section $\sigma(\xi,r)$ is
universal, only $|\Psi_{T,L}|^{2}$ depend on $Q^{2}$ and $m_{f}^{2}$.

In [4-6] we have shown that $\sigma(\xi,r)$ satisfies the
generalized BFKL equation
\beq
{\partial \sigma(\xi,r) \over \partial \xi} ={\cal K}\otimes
\sigma(\xi,r)\, ,
\label{eq:2.5}
\endeq
where in terms of the expansion (\ref{eq:2.1})
the kernel ${\cal K}$ is defined by
\arr
\sigma_{n+1}(r)={\cal K}\otimes\sigma_{n}(r)=~~~~~~~~~~~~~~~~~~~~~~~~~~~~
\nonumber\\
{3 \over 8\pi^{3}} \int d^{2}\vec{\rho}_{1}\,\,
\mu_{G}^{2}
\left|g_{S}(R_{1})
K_{1}(\mu_{G}\rho_{1}){\vec{\rho}_{1}\over \rho_{1}}
-g_{S}(R_{2})
K_{1}(\mu_{G}\rho_{2}){\vec{\rho}_{2} \over \rho_{2}}\right|^{2}
[\sigma_{n}(\rho_{1})+
\sigma_{n}(\rho_{2})-\sigma_{n}(r)]   \, \, .
\label{eq:2.6}
\endarr
Here $R_{c}=1/\mu_{G}$ is the correlation radius for perturbative
gluons, $R_{i}={\rm min}\{r,\rho_{i}\}$,
$g_{S}(r)$ is the effective colour charge,
\beq
\alpha_{S}(r)={g_{S}(r)^{2}\over 4\pi}={4\pi \over
\beta_{0}\log\left(
{C^{2} \over \Lambda_{QCD}^{2}r^{2}}\right)} \, ,
\label{eq:2.7}
\endeq
where $C\approx 1.5$ [2] and we impose the infrared
freezing $\alpha_{S}(r>R_{f})=\alpha_{S}^{(fr)}=0.8$ [6].

In the BFKL scaling limit of
$r,\rho_{1} \ll R_{c}$ and of fixed $\alpha_{S}$, the kernel
${\cal K}$ becomes independent of $R_{c}$, Eq.~(\ref{eq:2.5})
becomes equivalent to the BFKL equation [1] and has the familiar
BFKL eigenfunctions
\beq
E(\xi,r)=r^{1+2i\nu}
\exp[\xi\Delta(i\nu)]
\label{eq:2.8}
\endeq
with the BFKL eigenvalue (intercept)
\arr
\Delta(i\nu) =
{3\alpha_{S} \over \pi} [2\Psi(1)-
\Psi({1\over 2}-i\nu)-\Psi({1\over 2}+i\nu)]\, ,
\label{eq:2.9}
\endarr
where $\Psi(x)$ is the digamma function.

%----------------------------------------------------

%---------------------   Section 3

%\section{The generalized BFKL equation
%and the DLLA to GLDAP equation}

Deep inelastic scattering at large $Q^{2}$ probes $\sigma(\xi,r)$
at small $r^{2} \propto 1/Q^{2}$ ([2-4], see also below).
In the DLLA of large but finite $\xi$ and large
$L(r)$,
the kernel ${\cal K}$ is dominated by
$r^{2}\ll \rho_{i}^{2}\ll R_{f}^{2}$, one can neglect
$\sigma(r)\ll \sigma(\rho_{i})$ and factor out
$\alpha_{S}(r)$ in Eq.~(\ref{eq:2.6}), which in the DLLA
takes on a particularly simple form [4-6]
\beq
\sigma_{n+1}(r)={\cal K}\otimes \sigma_{n}(r) =
{3 r^{2}
\alpha_{S}(r)
\over \pi^{2}}\int_{r^{2}}^{R_{f}^{2}}{d^{2}\vec{\rho}
\over \rho^{4}}
\sigma_{n}(\rho)
\, ,
\label{eq:3.1}
\endeq
what is equivalent to the large-${1\over x}$ limit of the
GLDAP evolution equation [7].
As a boundary conditon, one can start with the dipole cross section
for interaction with the nucleon target [2]
\beq
\sigma_{0}(r)=
{32 \over 9}
\int {d^{2}\vec{k}
\over(\vec{k}^{2}+\mu_{G}^{2})^{2} }
\alpha_{S}(k^{2})\alpha_{S}(\kappa^{2})
[1-G_{p}(3\vec{k}^{2})]
\left[1-\exp(-i\vec{k}\vec{r})\right] \, ,
\label{eq:3.2}
\endeq
where $G_{p}(q^{2})$ is the charge form factor of the proton,
$\kappa^{2}={\rm max}\{k^{2},{C^{2}\over r^{2}}\}$.
Then, iterations of Eq.~(\ref{eq:3.1}) produce the familar DLLA
solution (\ref{eq:1.3}) [4].

The question of whether there is a strong, experimentally
observable, difference between the BFKL and GLDAP evolutions
(the latter usually being considered to DLLA),
is being discussed in the literature for quite a time ([8,9]
and references therein). A comparison of the DLLA iterations
$
\sigma_{n+1}(r)={\cal K}\otimes \sigma_{n}(r) =
12L(r)\sigma_{n}(r)/(n+1)\beta_{0}$
with the BFKL iterations $\sigma_{n+1}(r)=
{\cal K}\otimes \sigma_{n}(r)=
\Delta_{\Pom}
\sigma_{n}(r)$
suggests the DLLA breaking at $L(r)/n \approx L(r)/\sqrt{\eta} \lsim
{3\over 4} \Delta_{\Pom}$.
(At large $\eta$ the dominant contribution to
the DLLA cross section (\ref{eq:1.3}) comes from
$n \sim \sqrt{\eta}$.)
We can go one step further and compare in detail
the $\xi$-dependence of the
solution of our BFKL equation and of the DLLA solution of the
GLDAP equation,
starting with the identical initial condition Eq.~(\ref{eq:3.2}).
(For the sake of definitness we consider $R_{c}=0.275$f which
gives $\Delta_{\Pom}=0.4$ [6].)
Specifically, we compare the effective intercepts
$\Delta_{eff}(\xi,r)=\partial \log\sigma(\xi,r)/\partial \xi$ for
the two solutions.
Since the DLLA  asymptotics
(\ref{eq:1.3}) works at $\xi \gsim 1$, we shall consider the
cross section (\ref{eq:3.2}) as a result of the GLDAP evolution
by $\xi_{0}$ units from a lower energy, i.e., we take for the
DLLA  solution
\beq
\sigma_{DLLA}(\xi,r)=\sigma_{0}(r)
\sqrt{{ \xi_{0}\over\xi_{0}+\xi}}\exp\left[
2\sqrt{{4\over 3}\L(r)}\left(\sqrt{\xi_{0}+\xi}-\sqrt{\xi_{0}}
\right)\right] \,.
\label{eq:3.5}
\endeq
It satisfies $\sigma_{DLLA}(\xi=0,r)=\sigma_{0}(r)$ by the
construction, and gives the DLLA effective intercept
\beq
\Delta_{DDLA}(\xi,r)=
\sqrt{{4L(r)\over 3(\xi_{0}+\xi)} }-
{1\over 2(\xi_{0}+\xi)} \, .
\label{eq:3.6}
\endeq
The same $\sigma_{0}(r)$ is taken as the
boundary condition for the BFKL equation (\ref{eq:2.5}). (Here
we assume it to correspond to $x=x_{0}\approx
3\cdot 10^{-2}$, the more detailed BFKL phenomenology of deep
inelastic scattering will be presented elsewhere [10]).

Firstly we compute
$\Delta_{eff}(\xi=0,r)$ from our BFKL equation (\ref{eq:2.5}) and
make the readjustment
\beq
L(r) \Longrightarrow \log\left[{\alpha_{S}^{(fr)}\over
\alpha_{S}(r)}\right] +c
\label{eq:3.7}
\endeq
such that
$\Delta_{DLLA}(\xi=0,r)$ of Eq.~(\ref{eq:3.6}) gives a good
approximation to this $\Delta_{eff}(\xi=0,r)$ at small $r$.
(Recall that $L(r)$ is defined up to an additive
constant $c\lsim 1$.)
With $\xi_{0}=1.25$ this is achieved taking $c\approx 0.05$.
Secondly, we study how the BFKL and DLLA
effective intercepts diverge at large $\xi$ (Fig.1).
At $\xi =\log({x_{0}\over x})\sim 1$,
both the DLLA and BFKL effective
intercepts are smaller than $\Delta_{\Pom}$ at $r\gsim 0.2$f
and are larger than $\Delta_{\Pom}=0.4$ at smaller $r$.
The good matching of the BFKL and DLLA effective intercepts at
small $r$ is not surprizing, since our generalized BFKL equation
(\ref{eq:2.5},\ref{eq:2.6}) has
the GLDAP equation as a limiting case at small $r$ [4-6].
With rising ${1\over x}$, the BFKL
effective intercept flattens and
tends to $\Delta_{\Pom}$, rising at large $r$
and decreasing at small $r$, whereas the DLLA
intercept monotonically decreases with $\xi$ at all $r$, until
the DLLA  breaking $\Delta_{DLLA}(\xi,r)\leq \Delta_{\Pom}$ takes
place at $\xi=\xi_{c}(r)$ given by
\beq
\xi_{c}(r)=\log\left({x_{0}\over x_{c}(r)}\right)=
{4\over 3\Delta_{\Pom}^{2}}\log\left[{1\over \alpha_{S}(r)}\right] \, .
\label{eq:3.8}
\endeq
(The much discussed boundary suggested in [11] is not born out by our
accurate comparison of the DLLA and BFKL solutions.)
The intercept $\Delta_{\Pom}$ is small,
$\Delta_{\Pom}=0.4$ at $\mu_{G}=0.75$GeV [6]. The resulting large
numerical factor
\beq
{4\over 3\Delta_{\Pom}^{2}} \approx 8 \, ,
\label{eq:3.9}
\endeq
which emerges in the r.h.s. of Eq.~(\ref{eq:3.8}), explains why
the close similarity of the BFKL and DLLA solutions persists in
such a broad range of $r$ and $x$, relevant to the HERA experiments.

%----------------------------------------------

%-------------------      section 4

%\section{The DLLA identity and new solutions of the generalized
%BFKL equation}

The dipole cross section $\sigma(\xi,r)$ is related to the more
familiar gluon  structure function $G(\xi,r)$ by [4,12]
\beq
G(\xi,r)=xg(x,r)={3\sigma(\xi,r) \over \pi^{2}r^{2}\alpha_{S}(r)} \, ,
\label{eq:4.1}
\endeq
where $g(x,r)$ is the density of gluons at $x=x_{0}\exp(-\xi)$ and
the virtuality $Q^{2}\sim 1/r^{2}$. In terms of the $G(\xi,r)$
the DLLA equation (\ref{eq:3.1}) is equivalent to
\beq
\kappa(\xi,r) = {\beta_{0}\over 12}\cdot {1\over G(\xi,r)}
{\partial^{2} G(\xi,r)\over \partial \xi\,\partial L(r)} =1\, ,
\label{eq:4.2}
\endeq
which we shall refer to as the DLLA
identity. One can easily evaluate $\kappa(\xi,r)$ for the
experimentally measured gluon distributions.
It is interesting to look at the possible departure from the
DLLA identity of the above described solution of
our generalized BFKL equation (\ref{eq:2.5},\ref{eq:2.6})
subject to the boundary condition (\ref{eq:3.2}).
The results of such a test are shown in Fig.2.  We find that our
BFKL solution produces $\kappa(\xi,r) \approx 1$ in a very broad
range of $\xi$ and $r$ of the practical interest:
the DLLA identity holds to the few per cent accuracy at
$r \lsim {1\over 3}R_{c}$, and to the
$(20-30)\%$ accuracy even at large $r$, up to $r^{2} \lsim
{1\over 2}R_{c}^{2}$. The somewhat oscillatory $r$-dependence
of the $\Delta_{eff}(\xi,r)$ is quite natural and has its
origin in the contribution of oscillating harmonics with large
$|\nu|$ (for instance, see Eq.~(\ref{eq:2.8})), which
die out at large $\xi$.

This remarkable finding of $\kappa(\xi,r)\approx 1$ can be
understood as follows:
In the weak coupling limit
we can factor out $\alpha_{S}(r)$ from
the kernel ${\cal K}$, and
the generalized BFKL equation (\ref{eq:2.5},\ref{eq:2.6})
for $G(\xi,r)$ takes the form
\beq
{\partial G(\xi,r) \over \partial \xi}=
{3\over 2\pi^{2}}\int d^{2}\vec{\rho}_{1}\left[
{\alpha_{S}(\rho_{1})G(\xi,\rho_{1})\over \rho_{2}^{2}}+
{\alpha_{S}(\rho_{2})G(\xi,\rho_{2})\over \rho_{1}^{2}}-
{r^{2}\alpha_{S}(r)G(\xi,r)\over \rho_{1}^{2}\rho_{2}^{2}}\right]\,,
\label{eq:4.3}
\endeq
in which the leading contribution comes from
$\rho_{i}^{2} \gsim r^{2}$. The major function of
the term $\propto \alpha_{S}(r)G(\xi,r)$ is a regularization
of the logarithmic singularity at $\rho_{1,2}\rightarrow 0$,
and at $\alpha_{S}(r) \ll 1$ this term can be neglected at
the expense of the integration cutoff $\rho_{1,2}^{2}\gsim r^{2}$.
As a result,
Eq.~(\ref{eq:4.3}) takes the form which is identical to
the GLDAP equation (\ref{eq:3.1}):
\beq
{\partial G(\xi,r) \over \partial \xi}=
{12\over \beta_{0}}\int_{0}^{L(r)}dL(\rho)\,\,
G(\xi,r)\,.
\label{eq:4.4}
\endeq
Now we wish to emphasize that making DLLA is not imperative
when solving the GLDAP equation, and
evidently
Eq.~(\ref{eq:4.4}) has the one-parametric family of the small-$r$
eigenfunctions
\beq
G(\gamma,\xi,r)=\exp[\gamma L(r)]\exp(\Delta\xi)=
\left[{1\over \alpha_{S}(r)}\right]^{\gamma}\exp(\Delta\xi)\, ,
\label{eq:4.5}
\endeq
where the exponent $\gamma$ is related to the intercept $\Delta$ by
Eq.~(\ref{eq:1.2}). The corresponding dipole
cross section is given by Eq.~(\ref{eq:1.1}).
In contrast to the conventional DLLA solution (\ref{eq:4.5}) which
only summs the leading powers $[L(r)\xi]^{n}$, our new solution
(\ref{eq:4.5},\ref{eq:1.1}) manifestly
summs all powers $L(r)^{k}$. One can easily check that
the neglected $\propto \alpha_{S}(r)G(\xi,r)$ term in
Eq.~(\ref{eq:4.3}) gives the $\propto \alpha_{S}(r)$
correction to the solution (\ref{eq:4.5}).
The small-$r$ considerations alone do not fix the intercept
$\Delta$ and the exponent $\gamma$, they are determined from the
matching the solution (\ref{eq:1.1}) with the large-$r$
solution of our generalized BFKL equation. As we have shown in [6],
the value of $\Delta_{\Pom}$ is predominatly controlled by
the semiperturbative region of $r\sim R_{c}$.

In Fig.3a we show the pomeron dipole section
found in [6] by a numerical solution of our generalized BFKL
equation. In Fig.3b we show that these solutions to a good
accuracy satisfy the property
\beq
\chi={\sigma_{\Pom}(r)\over r^{2}} \alpha_{S}(r)
^{\gamma-1} =const \,.
\label{eq:4.6}
\endeq
Typically, this property holds up to $r\lsim {1\over 2}R_{c}$.
The smaller is the gluon correlation radius $R_{c}$, the sooner
starts, and the larger becomes with the increasing $r$,
the $\propto \alpha_{S}(r)$ correction to Eq.~(\ref{eq:1.1})
(we keep $\alpha_{S}^{(fr)}=0.8$).

When solving the GLDAP equations exactly (numerically), one
of course implicitly summs all
powers of $\log[1/\alpha_{S}(Q^{2})]$.
The obvious conclusion from the above derivation of
(\ref{eq:4.5},\ref{eq:1.1}) is that the DLLA solution
(\ref{eq:1.3}), which only is valid at moderate values
of $\xi$, evolves at larger $\xi$, beyond the boundary (\ref{eq:3.8}),
into our new solution
(\ref{eq:1.1}). Consequently, from the point of view of the
practical phenomenology, there is no real clash between the GLDAP and
BFKL evolutions. The BFKL evolution is evolution in $\xi$ and requires
as the boundary condition the knowledge of
$\sigma(\xi=0,r)$ for all $r$ at fixed $\xi=0$. The GLDAP evolution
is evolution in $L(r)$, and requires the knowledge of
$\sigma(\xi,r_{0})$ for all $\xi$ at fixed $r=r_{0}$.
Inspection of Fig.~1 shows that
at
\beq
r=r_{0}\sim (0.3-0.7)R_{c}\,,
\label{eq:4.7}
\endeq
the DLLA and the BFKL intercepts
are very close to each other and to the $\Delta_{\Pom}$ in a broad
range of $x$ of the interest for the HERA experiments.
Our conclusion is that choosing for the GLDAP evolution
the boundary condition at $r_{0}\sim 0.15$f, i.e., at
$Q_{0}^{2}\sim$ 10-20GeV$^{2}$ (for the $Q^{2}-r^{2}$ relationship
see below),
\beq
G(x,Q_{0}^{2}),\,xq(x,Q_{0}^{2}),\,x\bar{q}(x,Q_{0}^{2}) \propto
x^{\Delta} \, ,
\label{eq:4.8}
\endeq
one will obtain the GLDAP solutions, which for all the practical
purposes will be indistinguishable from the BFKL solution, if
$\Delta$ is taken equal to $\Delta_{\Pom}$.

%------------------------------------------------------

%---------------------------   Section 5

%\section{Universal structure function in the BFKL regime}

Making use of properties of the modified  Bessel functions,
after the $z$-integration one can write
\arr
\sigma_{T}(\gamma^{*}N,\xi,Q^{2})=
\int_{0}^{1} dz\int d^{2}\vec{r}\,\,
\vert\Psi_{T}(z,r)\vert^{2}\sigma(\xi,r)\nonumber\\
\propto
{1\over Q^{2}} \int_{1/Q^{2}}^{1/m_{f}^{2}}
{d r^{2} \over r^{2} }{\sigma(\xi,r)\over r^{2}} \propto
{1\over Q^{2}}\int_{0}^{-\log\alpha_{S}(Q^{2})}
dL(\rho)\,\, G(\xi,\rho)
\label{eq:5.1}
\endarr
Notice, that the factor $1/Q^{2}$ in Eq.~(\ref{eq:5.1}),
which provides the Bjorken scaling,  comes
from the probability of having the $q\bar{q}$ fluctuation
of the highly virtual photon. Making use of the pomeron
soluiton (\ref{eq:4.5}) in (\ref{eq:5.1}), we easily
obtain the universal asymptotic
structure function (\ref{eq:1.6}).
Since $G(\xi,r)$ slowly rises towards small $r$, the structure
function $F_{2}(x,Q^{2})$ receives a substantial (but not quite
dominant) contribution from
\beq
r^{2} \sim {B\over Q^{2}} \, .
\label{eq:5.3}
\endeq
The explicit form of the wave function (\ref{eq:2.3}) leads to
a rather large numerical factor $B\sim 10$ in the relationship
(\ref{eq:5.3}) (for the related slow onset of the short-distance
dominance also see [13]). According to Fig.~1, the BFKL effects
are most significant at $r\gsim {1\over 2}R_{c}\approx 0.15$f,
and the scaling violations in the region of
$Q^{2}\lsim$(20-30)GeV$^{2}$ seem to be the most interesting ones
from the point of view of testing the onset of the BFKL regime.

Eq.~(\ref{eq:1.6}) predicts the universal, $x$-independent,
scaling violation at small $x$:
\beq
-{\partial \log F_{2}(x,Q^{2}) \over \partial \log\alpha_{S}(Q^{2})}
=  {4\pi \over \beta_{0}\alpha_{S}(Q^{2})}
{\partial \log F_{2}(x,Q^{2}) \over \partial \log Q^{2}}
= \gamma                                            \, .
\label{eq:5.2}
\endeq
The exponent $\gamma$ and the pomeron intercept $\Delta_{\Pom}$
are not calculable within the perturbative QCD. For the guidance,
we cite the results of an analysis [6]: $\gamma=2.19,\,2.81,\,
3.33,\, 3.70$ for $\mu_{G}=0.3,\,0.5,\,0.75,\,1.0$ GeV,
respectively. It would have been of great interest to see the
universal scaling violation (\ref{eq:5.2}) experimentally,
but closer inspection of Fig.~1 makes us to conclude that
the applicability region of (\ref{eq:5.2}) lies somewhat beyond
the kinematical range of the HERA experiments. Indeed, the universal
scaling violation comes along with the  $r$-independent universal
$x$-dependence of the structure function (\ref{eq:1.6}), i.e, with
$\Delta_{eff}(\xi,r)=\Delta_{\Pom}$, whereas Fig.~1 shows that in the
kinematic range of HERA, the $\Delta_{eff}(\xi,r)$ for our BFKL
solution exhibits a still substantial $r$-dependence.

%---------------------------------------------------

% ---------------------    Section 6

%\section{The running versus fixed coupling $\alpha_{S}(r)$
%for the pomeron cross section}

The discussion of the BFKL effects in the current literature
concentrates upon the approximation of fixed $\alpha_{S}$
(for the review and references see [8]). Our finding is that
the effect of the running coupling constant is quite substantial.
Firstly, the pomeron cross section (\ref{eq:1.1}) dramatically
differs from the $\propto r^{1}$ BFKL scaling solution
$(\ref{eq:1.5},\ref{eq:2.8})$ for the fixed $\alpha_{S}$. Secondly,
one can define the counterpart of the DLLA identity
for fixed $\alpha_{S}$ too, making the straightforward substitution
\beq
L(r)={\beta_{0} \over 4\pi}\int_{r^{2}}^{R^{2}}
{d\rho^{2}\over \rho^{2}}\alpha_{S}(\rho)\Longrightarrow
{\beta_{0} \over 4\pi}\alpha_{S}\log\left({R^{2}\over r^{2}}\right)\,.
\label{eq:6.1}
\endeq
Then, making use of the BFKL formula for the pomeron intercept [1]
\beq
\Delta_{\Pom} = \Delta(0)={12\log2 \over \pi}\alpha_{S} \, ,
\label{eq:6.2}
\endeq
one readily finds $\kappa=2\log2$
for the BFKL pomeron solution (\ref{eq:1.5}).
This departure from $\kappa=1$ for the case of the running
$\alpha_{S}(r)$ also emphasizes a dramatic difference between the
cases of the fixed and running strong coupling.
Therefore, the fixed-$\alpha_{S}$ considerations are not
appropriate for the phenomenology of deep-inelastic scattering.

\section{Conclusions}

The purpose of this paper has been a comparison of the BFKL and
GLDAP evolutions at large ${1\over x}$
 in the framework of the dipole cross
section description of deep inelastic scattering. We found
a new solution (\ref{eq:1.1}) which is common to
the BFKL equation with
the running QCD coupling  and to the GLDAP equation, and provides
a smooth matching of our generalized BFKL evolution
and of the GLDAP evolution
at small $x$. We derived the asymptotic form of the structure
function (\ref{eq:1.6}) and the universal law
Eq.~(\ref{eq:5.2}) for the scaling violations.
Our principle conclusion is that the GLDAP
evolution remains a viable phenomenology of the scaling violations
at HERA and beyond, provided that one starts with the boundary condition
(\ref{eq:4.8}) at $Q_{0}^{2}\sim$(10-20)\,GeV$^{2}$.
\bigskip\\
{\bf Acknowledgements}: B.G.Z. is grateful to
J.Speth for the hospitality at IKP, KFA J\"ulich, where this
work was initiated.
\pagebreak

{\bf Figure captions:}

\begin{itemize}
\item[Fig.1 - ]
Comparison of effective intercepts of the DLLA solution
(\ref{eq:3.5}) and of the solution of
our BFKL equation. Both solutions start with the identical dipole
cross section at $x=3\cdot 10^{-2}$. The pomeron intercept
$\Delta_{\Pom}=0.4$ is shown by the horizontal line.

\item[Fig.2 - ]
Test of the DLLA identity for the solution of our generalized
BFKL equation ($R_{c}=0.27$f, \,$\mu_{G}=0.75$GeV).

\item[Fig.3 - ]
The left box:
the pomeron dipole cross section $\sigma_{\Pom}(r)$ for different
values of $\mu_{G}$. The straight lines show the $r^{1}$ and $r^{2}$
behavior.\\
The right box: test of the scaling law $\chi=
{1\over r^{2}}\sigma_{\Pom}(r)[\alpha_{S}(r)]
^{\gamma-1}$
=const.
\end{itemize}

\begin{thebibliography}{299}

\bibitem{1}
E.A.Kuraev, L.N.Lipatov and V.S.Fadin, {\sl Sov.Phys. JETP}
{\bf 44} (1976) 443; {\bf 45} (1977) 199;
Ya.Ya.Balitskii and L.N.Lipatov, {\sl Sov. J. Nucl. Phys.}
{\bf 28} (1978) 822;
L.N.Lipatov, {\sl Sov. Phys. JETP} {\bf 63} (1986) 904;
L.N.Lipatov. Pomeron in Quantum Chromodynamics. In: {\sl Perturbative
Quantum Chromodynamics}, editor A.H.Mueller, World Scientific, 1989.

\bibitem{2}
N.N.~Nikolaev and B.G.~Zakharov, {\it Z. Phys.} {\bf C49} (1991) 607;
{\bf C53} (1992) 331.

\bibitem{3}
V.Barone, M.Genovese, N.N.Nikolaev, E.Predazzi and B.G.Zakharov,
{\sl Z. Phys.} {\bf C58} (1993) 541.

\bibitem{4}
N.N.Nikolaev and B.G.Zakharov, The triple-pomeron regime and the
structure function of the pomeron in the diffractive deep inelastic
scattering at very small $x$,
{\sl Landau Inst. preprint}
{\bf Landau-16/93} and {\sl J\"ulich preprint}
{\bf KFA-IKP(Th)-1993-17}, June 1993, submitted to
{\sl Z. Phys.} {\bf C}.

\bibitem{5}
N.N.Nikolaev, B.G.Zakharov and V.R.Zoller, The $s$-channel
approach to Lipatov's pomeron and hadronic cross sections,
{\sl J\"ulich preprint}
{\bf KFA-IKP(TH)-1993-34}, December 1993; {\sl JETP Letters},
to be published.

\bibitem{6}
N.N.Nikolaev, B.G.Zakharov and V.R.Zoller, The spectrum and
solutions of the generalized BFKL equation for total cross section,
{\sl J\"ulich preprint}
{\bf KFA-IKP(TH)-1994-1}, January 1994; submitted to
{\sl Phys. Lett.} {\bf B}.

\bibitem{7}
V.N.~Gribov and L.N.~Lipatov, {\it Sov. J. Nucl. Phys.} {\bf 15} (1972)
438; L.N.~Lipatov, {\it Sov. J. Nucl. Phys.} {\bf 20} (1974) 181;
Yu.L.~Dokshitser, {\it Sov. Phys. JETP} {\bf 46} (1977) 641;
G.~Altarelli and G.~Parisi, {\it Nucl. Phys.} {\bf B126} (1977) 298.

\bibitem{8}
B.Badelek, K.Charchula, M.Krawczyk
and J.Kwiecinski, {\sl Rev. Mod. Phys. } {\bf 64} (1992) 927.

\bibitem{9}
G.Marchesini and B.R.Webber, {\sl Nucl. Phys.} {\bf B349} (1991) 617.

\bibitem{10}
N.N.Nikolaev and B.G.Zakharov, Deep inelastic scattering and the
BFKL pomeron,
paper in preparation.

\bibitem{11}
L.V.Gribov, E.M.Levin and M.G.Ryskin, {\sl Phys.~Rep.} {\bf 100}
(1983) 1.

\bibitem{12}
V.~Barone, M.Genovese,
N.N.~Nikolaev, E.~Predazzi and B.G.~Zakharov, Unitarization of
structure functions at large ${1\over x}$,
{\sl Torino preprint} {\bf DFTT 28/93}, June 1993, submitted
to {\sl Phys.Lett.} {\bf B}.

\bibitem{13}
B.Z.Kopeliovich, J.Nemchik, N.N:Nikolaev and B.G.Zakharov,
{\sl Phys. Lett.} {\bf B309} (1993) 179;
Deciive test of colour transparency in exclusive electroproduction
of vector mesons, {\sl J\"ulich preprint} {\bf KFA-IKP(TH)-1993-29},
{\sl Phys. Lett.} {\bf B}, to be published.
\pagebreak\\
\end{thebibliography}
\end{document}